# High-frequency magnetotransport in $LaMnO_3$ samples synthesized by microwave irradiation versus conventional heating

**Y. H. Lee, M. Manikandan, and R. Mahendiran***

Department of Physics, 2 Science Drive 3, National University of Singapore, Singapore-117551

**Corresponding author E-mail: phyrm@nus.edu.sg*

**Abstract**

Magnetoresistance of hole-doped $LaMnO_3$ at frequencies above a few MHz has been seldom reported compared to numerous studies on magnetoresistance measured at direct current. Here, we contrast the high-frequency magnetoresistance in the frequency range $f$ = 0.9-3 GHz in polycrystalline $LaMnO_3$ (LMO) synthesized by microwave irradiation of oxide precursors in a microwave furnace (MW-LMO) and by conventional heating (CH-LMO) in an electrical furnace. The structure at room temperature changed from orthorhombic in CH-LMO to rhombohedral in MW-LMO. A combination of magnetization and resistivity studies suggest that the CH-LMO sample is a canted antiferromagnetic insulator below 140 K but the MW-LMO is a ferromagnetic metal with $T_C$ = 240 K. While the high-frequency resistance of the CH-LMO sample at 300 K decreases gradually with increasing strength of the applied *dc* magnetic field for all $f$, a peak appears at $H = H_r > 0$ when $f \geq 1.4$ GHz in the MW-LMO sample and $H_r$ increases linearly with $f$. We attribute the observed features in the high-frequency magnetoresistance of MW-LMO to current-driven resonant excitation of spins in the paramagnetic state. Its absence in the CH-LMO sample highlights the need for an optimum hole density to observe this effect.

## I. INTRODUCTION

Hole-doped Mn-perovskites of the formula $La_{1-x}A_xMnO_3$ (A = Sr, Ca, Ba, etc.) have been extensively studied over the past three decades for their prospects in technological applications and in pursuit of understanding the mechanisms of colossal negative magnetoresistance exhibited by them. [ 1 ]. However, the vast majority of available studies deal with magnetoresistance measured using a direct current (*dc*). Reports on alternating current (*ac*) resistivity or magnetoresistance above a few kHz frequencies are very few [2,3,4,5,6,7]. The *ac* electrical transport with or without an external magnetic field in the MHz frequency range is interesting due to the following reasons: 1. It can detect the onset of ferromagnetism in low-resistivity samples without using an external magnetic field or a mutual inductance coil. For example, the *ac* resistance of a polycrystalline $La_{0.7}Sr_{0.3}MnO_3$ at $f$ = 5 MHz showed a sharp increase at the onset of ferromagnetic transition even in zero external magnetic field, whereas the *dc* resistance showed only a change in slope [8]. 2. Huge *ac* magnetoresistance (≈30-40%) was realized in low static magnetic fields ($H \leq 1$ kOe), whereas $H \geq 30$ kOe is generally needed to realize a similar value of magnetoresistance using a *dc* current [8]. 3. The *ac* electrical resistivity detected magnetic anomalies within the ferromagnetic state due to spin-reorientation in Sm doped $La_{0.7}Sr_{0.3}MnO_3$ [9] and also a structural transition in $Pr_{0.6}Sr_{0.4}MnO_3$ [10] although the *dc* resistivity showed no features. In essence, the temperature dependence of the *ac* resistivity in zero field mimicked that of *ac* magnetic permeability. No theoretical study is available yet to qualitatively model this phenomenon. Based on experiments, it was suggested that the complex impedance $Z = R + iX$ of a sample ($R$ is the *ac* resistance and $X$ is the reactance) was dependent on skin depth ($\delta = \sqrt{2\rho/\omega\mu_0\mu_t}$) which was influenced by transverse magnetic (*ac*) permeability ($\mu_t$), *dc* resistivity ($\rho$) and angular frequency ($\omega = 2\pi f$) of the *ac* current. A rapid increase of $\mu_t$ at the onset of ferromagnetic order in zero external magnetic field causes the skin depth to decrease that results in an abrupt increase of $R$ at $T_C$ in zero field. However, the application an external static magnetic field tends to suppress $\mu_t$ which in turn increases the skin depth and decreases *R* leading to colossal *ac* magnetoresistance in small *dc* magnetic fields for H= 200-1000 Oe .

An interesting question is what happens to the *ac* magnetoresistance when the frequency of the *ac* current is increased from MHz to GHz. Microwave absorption measured in $Nd_{0.7}Sr_{0.3}MnO_3$ single crystal by the cavity resonance method in the X- band frequency ($f \sim$ 9.5-10 GHz) indicated that the surface resistance of the sample was sensitive to small *dc* magnetic fields ($H$ < 1 kOe) [11]. However, magnetoresistance of an *ac* current-carrying manganite sample at GHz frequencies (different from the MW absorption measured using a resonant cavity) has been overlooked until recently. In our preliminary work on polycrystalline $La_{1-x}Ca_xMnO_3$ ($x$ = 0.35-0.45), we found that magnetoresistance at high frequencies ($f$ = 0.9-3 GHz) became positive and showed a peak for a specific *dc* field that shifted upward with the frequency of the *ac* current, however, these features disappeared outside this composition range at room temperature [12]. There is no report on the GHz magnetoresistance in the parent $LaMnO_3$ so far.

The purpose of this work is to compare high-frequency magnetotransport in the parent compound $LaMnO_3$ prepared by two different methods: 1. Microwave-assisted synthesis and as well as sintering in a microwave (MW) furnace in which chemical reactions between $La_2O_3$ and $Mn_2O_3$ are triggered by MW irradiation, and 2. Conventional heating (CH) of the reactants via a solid-state reaction using an electrical furnace, with a heat source consisting of high-current-carrying silicon carbide rods. Microwave-assisted synthesis is a novel method in which impinging microwave radiation is absorbed by reactants and converted into heat through intrinsic mechanisms [13,14,15,16]. Friction experienced by reorientation of electrical dipoles in response to oscillating MW electric field is considered to be the main mechanism of MW heating but oscillatory motion of ions/charges, eddy current, and magnetic losses can also contribute to additional heating [13,14]. To commence heating, at least one of the reactants must be a good microwave absorber and should exhibit high dielectric/magnetic loss at $f = 2.45$ GHz, which is the permitted frequency for commercial MW furnaces. The great advantage of the MW synthesis is that it takes ~ 1 - 30 minutes (depending on the MW absorbing capability of precursors) to raise the temperature of reactants from room temperature to 1000 °C and above, enabling rapid sintering of pellets, unlike the several days required in an electrical furnace. The internally generated heat diffuses from the interior to the surface of reactants (volumetric heating effect), whereas is from surface to interior in an electrical furnace. Beyond rapid heating, MW irradiation also allows lower sintering temperature and decrystallization in some materials, which require explanations beyond a simple thermal-driven effect [ 17]. Although MW irradiation has been exploited to synthesize $LaMO_3$ perovskite oxides ($M$ = Mn, Co, Fe, and Cr) using nitrates/hydroxides [13,18,19,20] of transition metals and rare earth ions, much remains to be known about how physical properties are modified by rapid heating and cooling involved in the MW synthesis compared with the synthesis by conventional heating (CH) in an electrical furnace. Hence, it is interesting to compare not only the high-frequency magnetoresistance but also the structure, magnetism, and *dc* resistivity in $LaMnO_3$ samples synthesized by the CH and MW methods.

## II. EXPERIMENTAL

Ten grams of $LaMnO_3$ (LMO) powders were prepared by mixing and grinding powders of preheated $La_2O_3$ and $Mn_2O_3$ taken in a stoichiometric ratio. Then, five grams of mixed LMO powder taken in an alumina crucible was placed in the middle of a multimode microwave muffle furnace (Milestone PYRO) and irradiated by microwave of frequency 2.45 GHz and power $P$ = 1.6 kW. The temperature of the crucible was monitored by an in-built infrared pyrometer. The temperature was programmed to reach 1100 ºC in 15 min (heating rate ~ 72 ºC/min) and dwelled at 1100 ºC for 20 min before switching off the MW power. The sample started cooling soon after the MW power was switched off and it cooled down to room temperature in 100 min (~ 12 ºC/min). The MW irradiated powder (MW-LMO) was again ground homogeneously and pelletized into a 10 mm diameter disc using a hydraulic press. The heating and cooling processes were repeated, but the sintering time at 1100 ºC was increased to 30 min. The obtained MW-LMO pellet was hard enough for the electrical/thermal and magnetic measurements. Another 5 grams of LMO powder was preheated at 1000 ºC for 12 hours in air using an electrical furnace and cooled to room temperature at a rate of 5 °C/min. Then, the powder was homogeneously ground and calcined at 1100 ºC for another 12 hours. Then, it was compressed into a 10 mm diameter pellet and sintered in the electrical furnace at 1100 ºC for 24 hours, followed by cooling to room temperature at a rate of 5 °C/min. The sample obtained via the conventional heating method is denoted as CH-LMO.

The structure and purity of the samples were verified using powder X-ray diffraction at room temperature. *Dc* magnetization was measured using a Lakeshore vibrating sample magnetometer from 400 K to 80 K and *dc* electrical resistivity was measured in a closed cycle refrigerator from 350 K to 80 K. A rectangular block of 2.2 mm x 3.4 mm x 1.7 mm size was cut from each of the above pellets for the *ac* magneto transport, coated with silver paint at the two ends and dried at 120 ºC for 5 min. The sample was loaded in a coaxial sample holder in which the sample bridges the core (signal line) and the ground plane, and the sample holder was placed between the pole pieces of an electromagnet. A radio-frequency impedance analyzer (Agilent E4991A) sources the MW current that flows from the signal line to the ground through Ag paint and the sample. The *ac* resistance $R$ and reactance $X$ of the sample were simultaneously recorded at multiple frequencies while the *dc* magnetic field generated by the electromagnet was slowly swept from $H$ = -2.6 kOe to $H$ = +2.6 kOe and back to +2.6 kOe.

The output power of the impedance analyzer was fixed at 1 mW to avoid heating. All the measurements were automated using the LabView$^{TM}$ graphical software. We define the *ac* magnetoresistance at a frequency $f$ as $MR$ (%) = 100 × [$R(H, f)$-$R(0, f)$]/R(0, $f$) and similarly for the magnetoreactance $MX$. To corroborate the results of the high-frequency magnetoresistance measurements, magnetic field-dependent microwave absorption of the sample was measured. The sample placed on a coplanar waveguide (NanOsc, Quantum Design Inc, USA) was excited by a microwave magnetic field arising from a signal line. The signal line of the CPW was driven by *ac* current ($f$ = 10 MHz to 7.8 GHz) from a vector network analyzer (Agilent, model N5230A) and the transmission coefficient ($S_{12}$) was recorded while a static magnetic field provided by an electromagnet was slowly swept.

## III. RESULTS

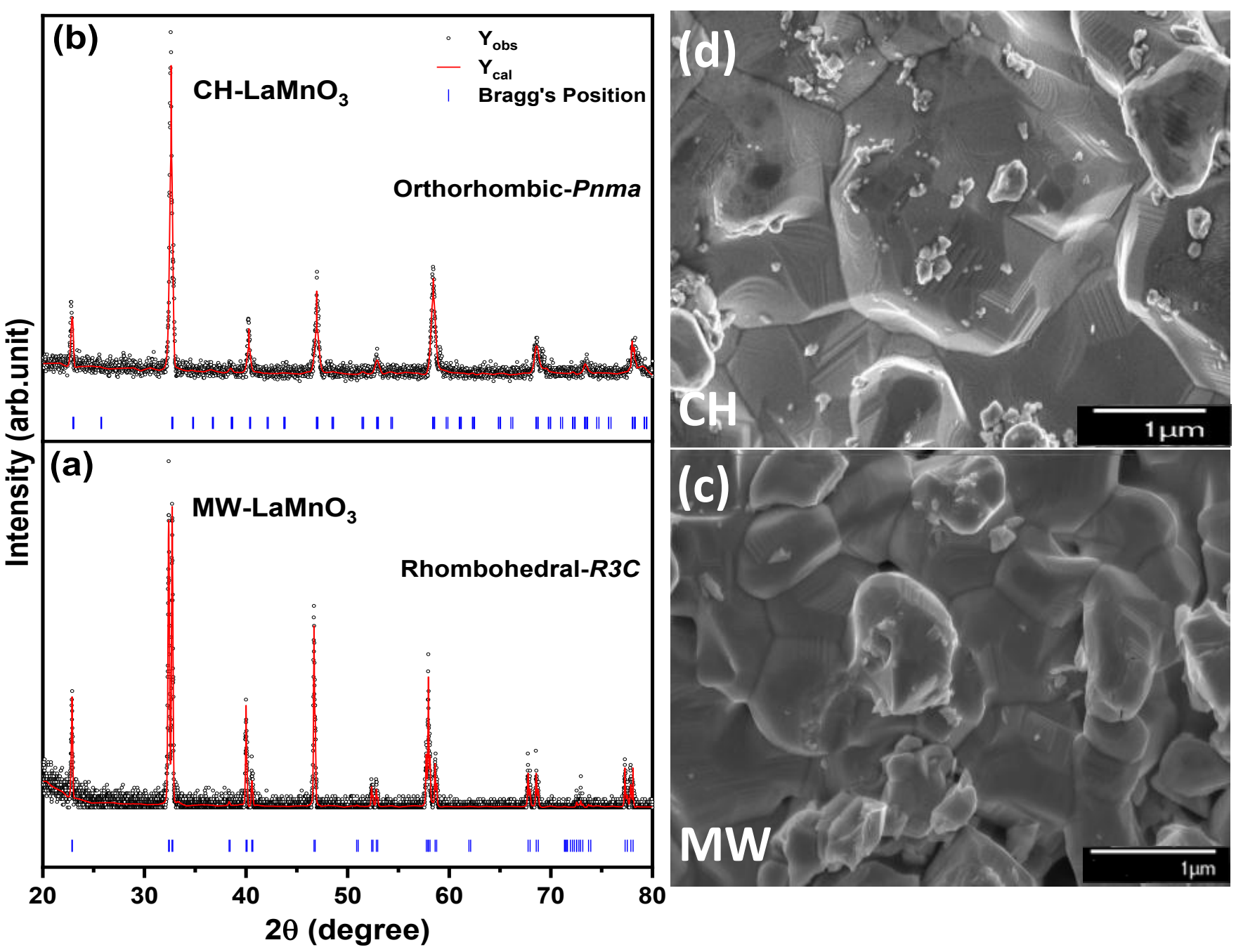

**FIG. 1.** Powder X-ray diffraction pattern (circles) and Rietveld fit for **(a)** MW-$LaMnO_3$ and **(b)** CH-$LaMnO_3$. Blue vertical lines show Bragg's positions. Scanning Electron Microscopy images of grains in **(c)** MW and **(d)** CH samples.

Figures. 1(a) & (b) show the powder X-ray diffraction patterns of the MW-LMO and CH-LMO samples, respectively along with their Rietveld refinement. Black open circles and red solid-line represent the recorded XRD patterns and Rietveld refinement fit, respectively. The MW-LMO crystallized in rhombohedral structure (space group *R-3c*) with lattice parameters $a = b = 5.527$Å, $c = 13.331$ Å and unit cell volume 352 $Å^3$, whereas the CH-LMO sample crystallized in the orthorhombic structure (space group *Pnma*) and the refined lattice parameters are $a = 5.470$ Å, $b = 5.473$ Å, $c = 7.734$ Å and the cell volume 231 $Å^3$. Fig. 1(c) & (d) compare the scanning electron microscopy (SEM) images of the MW-LMO and CH-LMO, respectively. Both samples exhibit similar morphologies of mixed dense and broken particles, but their sizes are larger in the CH-LMO. The average grain sizes are ~ 4 µm in the tCH-LMO and ~ 0.5 µm in the MW-LMO samples. The *ac* magnetotransport discussed later reflects the dynamic magnetic susceptibility of grains rather than a capacitive effect arising between semiconducting grain boundaries.

Fig. 2(a) compares the temperature-dependent *dc* magnetization, $M(T)$, of the CH-LMO and MW-LMO samples upon cooling from 400 K to 80 K under $H = 1$ kOe. The rapid increase of $M(T)$ around ~ 140 K in the CH-LMO reflects a change in the magnetic state from paramagnetism at high temperatures to canted antiferromagnetism at low temperatures. Stoichiometric $LaMnO_3$ is a layered antiferromagnet with a Neel temperature $T_N \approx 141$ K, however, canting of spins leads to ferromagnetic-like behavior in $M(T)$ in the field-cooled magnetization [21]. In contrast, the MW-LMO shows a ferromagnetic transition at $T = T_C = 240$ K. $M(H)$ isotherms shown in the two insets reveal that the MW-LMO has higher magnetization ($M = 3.15$ $\mu_B$/Mn at $H = 15$ kOe) than the CH-LMO sample ($M = 2.0$ $\mu_B$/Mn) at 80 K. Also, $M$ of MW-LMO is higher than that in the CH-LMO at 300 K, i.e. in the paramagnetic state. Fig. 2 (b) compares zero-field *dc* electrical resistivities ($\rho$) in both samples. While $\rho(T)$ of the CH-LMO is non-metallic down to 100 K (below which it is not measurable due to the instrument limit), the MW-LMO sample exhibits an insulator-metal transition with a peak near but below $T_C$.

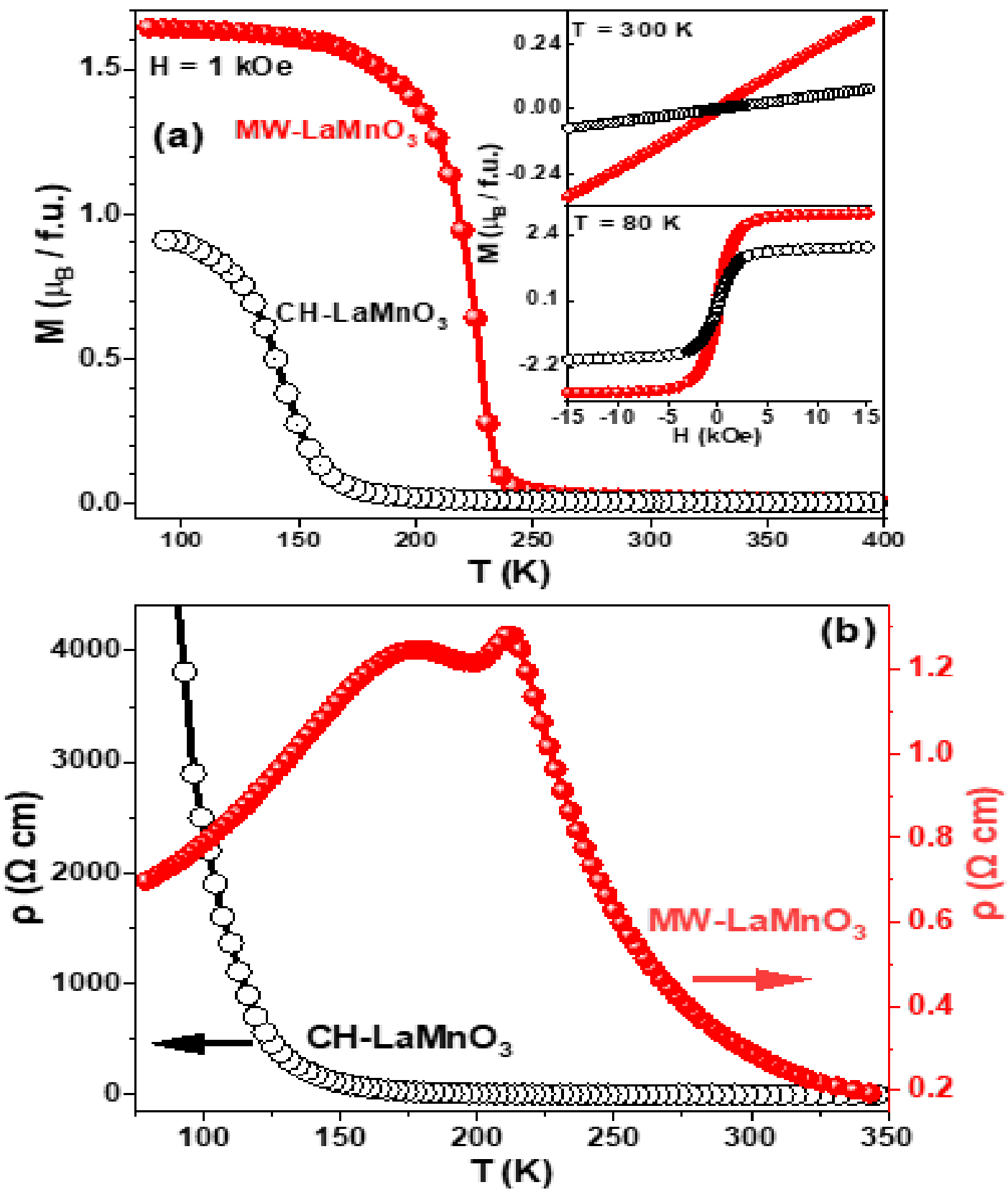

**FIG. 2. (a)** Temperature dependence of magnetization in the MW-LMO (red circle) and CH-LMO (black circle) upon cooling in a magnetic field of $H$ = 1 kOe. The insets compare the magnetization isotherms at 300 K and 80 K in both samples. **(b)** Temperature dependence of the *dc* resistivity in both the CH and MW-LMO samples in zero external magnetic field.

Fig. 3 shows magnetic field dependences of the *ac* resistance $R$ (left panel) and the reactance $X$ (right panel) of the CH-LMO for multiple frequencies ($f$ = 1 GHz to 3 GHz with 0.2 GHz step) when the dc magnetic field is swept between $H$ = 2.6 kOe and $H$ = -2.6 kOe. $R(H)$ decreases smoothly with increasing $H$ in both directions for all frequencies and the magnitude of R increases with increasing $f$. While the magnitude of $X(H)$ also increases with $f$, its field dependence is opposite to that of the resistance. It rises rapidly in low fields ( -0.5 kOe < $H$< 0.5 kOe) and tends to decrease at higher fields for $f$ < 1.4 GHz or to saturation at still higher frequencies.

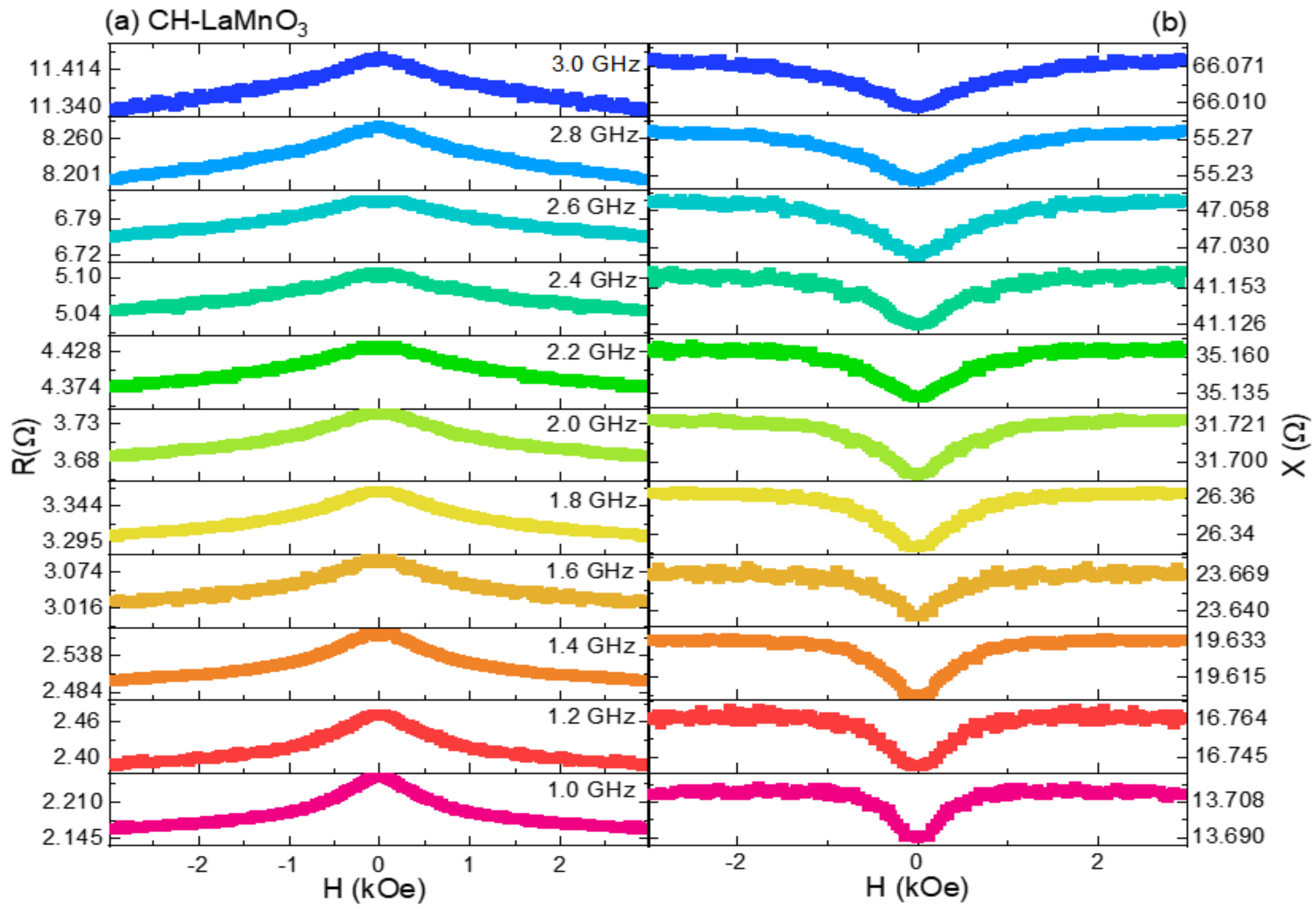

**FIG. 3.** Field dependence of **(a)** Resistance ($R$) and **(b)** reactance ($X$) of CH-LMO at selected frequencies, from $f$ = 1 GHz to 3 GHz at 300 K.

Fig. 4 illustrates $R(H)$ and $X(H)$ for the MW-LMO sample at 300 K. The field dependences of $R$ for $f$ = 1 and 1.2 GHz are similar to the CH-LMO sample. A shoulder develops in $R$ close to the zero field for either direction of the field when $f$ = 1.4 GHz, and the shoulder evolves into a prominent peak at 1.8 GHz. As $f$ increases further, the amplitude of this peak surpasses the feature at zero field and its position shifts towards higher fields. Also, the slope of $R(H)$ above the peak decreases with increasing frequency. $X(H)$ at 1 and 1.2 GHz are like the CH-LMO sample, but subtle changes in the field dependence of $X$ occur between 1.4

GHz and 3 GHz. A dip develops at a field away from the zero field and its position shifts towards a higher field with increasing frequency. The dip in $X(H)$ coincides with the zero crossing of $dR/dH$.

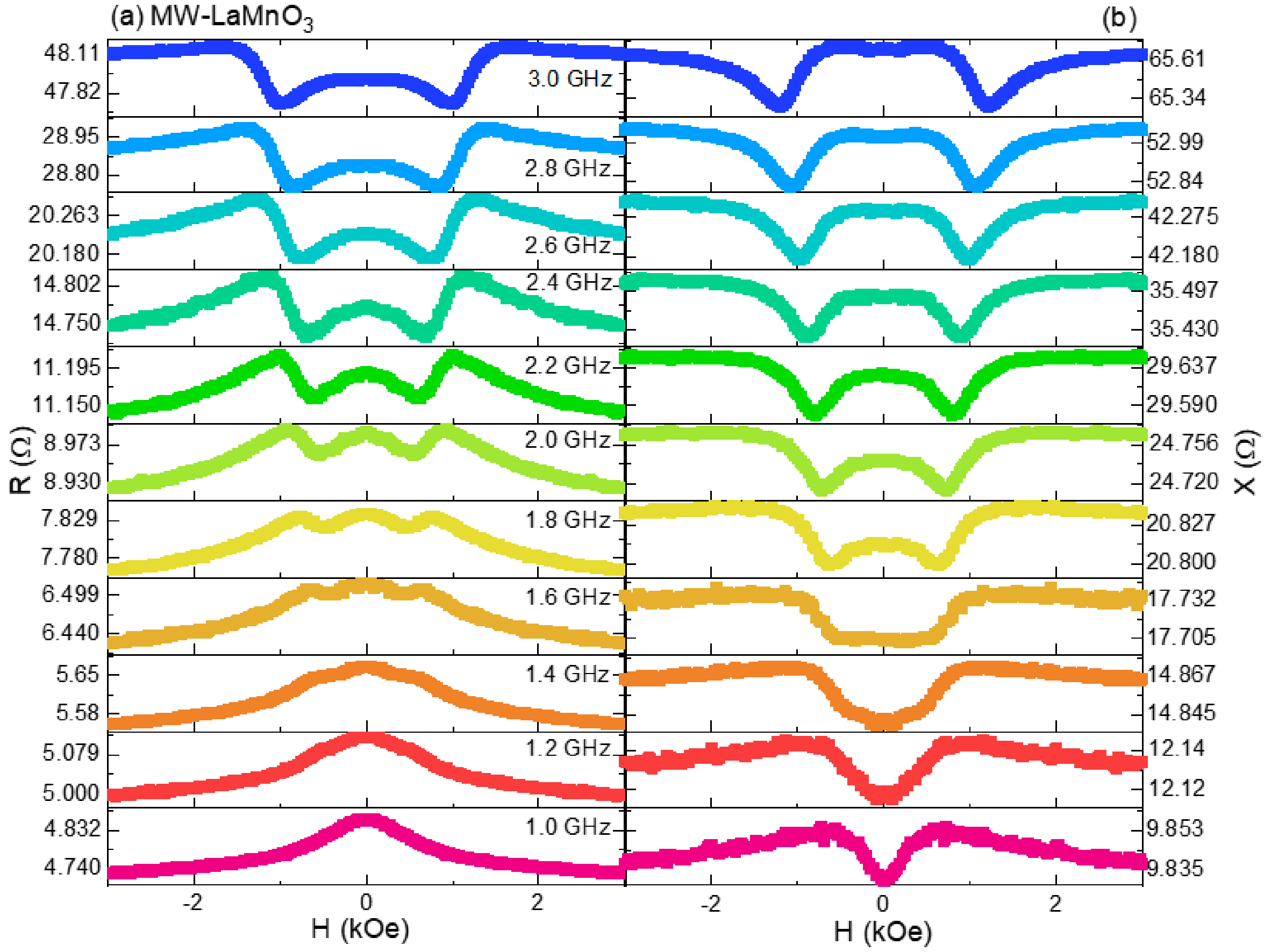

**FIG. 4.** Field dependence of **(a)** Resistance ($R$) and **(b)** reactance ($X$) of MW-LMO sample at selected frequencies from $f$=1 to 3 GHz at 300 K.

Fig. 5(a) shows the field dependence of the *ac MR* (%) for selected frequencies ($f$ = 2.2 – 3 GHz) for clarity. As the field increases from zero, the *MR* initially decreases, reaches a minimum (negative sign), then increases to a peak (positive sign), and finally decreases at higher fields. The magnitude of positive *MR* at the peak increases from 0.34 % at 2.2 GHz to 0.54 % at 3 GHz. We fit the *MR* using combinations of Lorentzian curves to extract the resonance field ($H_r$), which is plotted against the frequency in Fig. 5(b). It has been found that $H_r$ increases linearly as $f = \gamma H$.

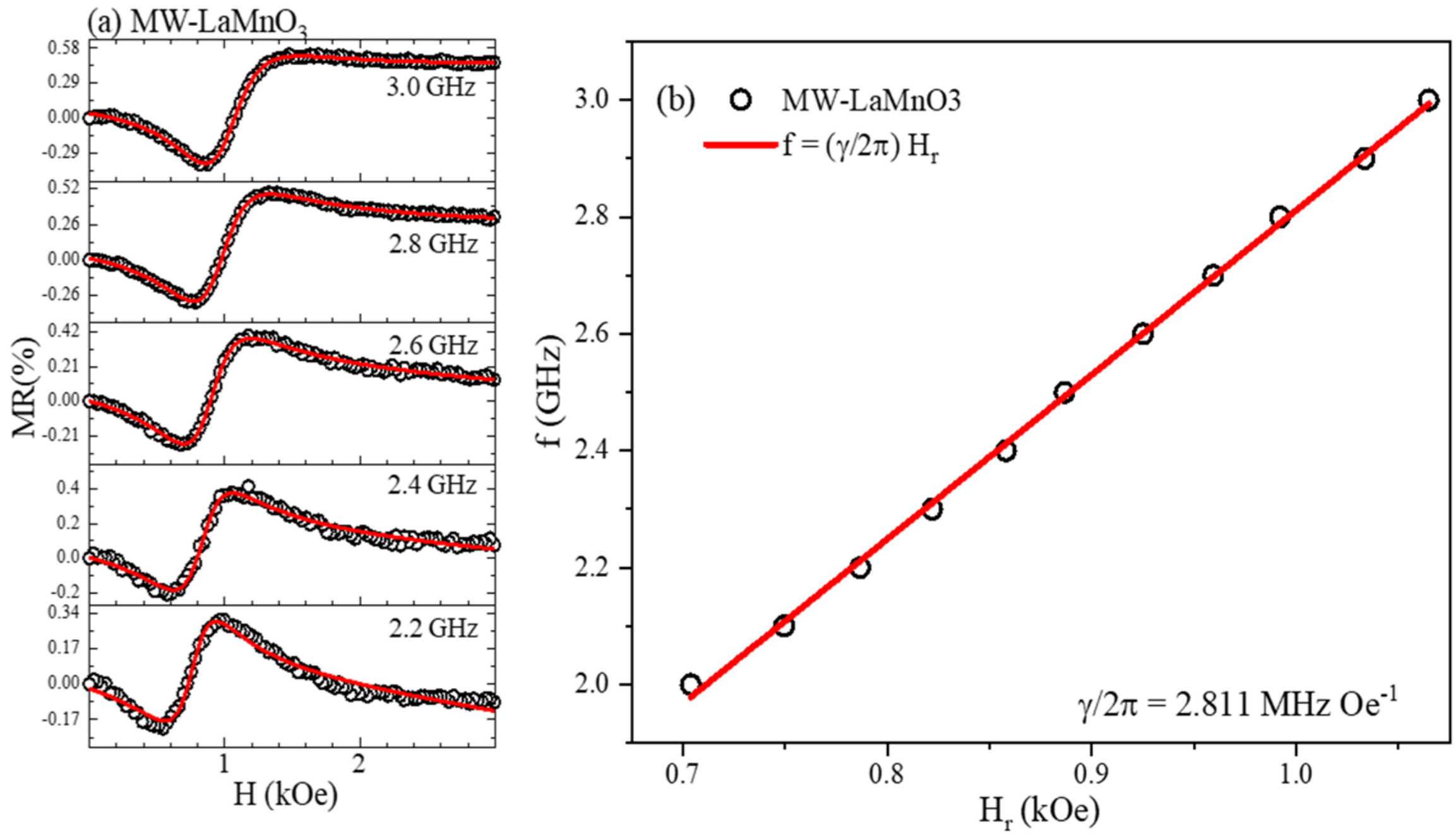

**FIG. 5. (a)** The field dependence of magnetoresistance (black open circles) of MW-LMO and the Lorentzian fit (red line) at different frequencies at 300 K. **(b)** Dependence of the resonance field ($H_r$) on the frequency (open circle) and the linear fit (red line) illustrating the linear relationship between $f$ and $H_r$ expected for electron spin resonance.

## IV. DISCUSSION

The $LaMnO_3$ sample synthesized by microwave heating shows distinct structural and physical properties: It has a rhombohedral structure and exhibits ferromagnetism, an insulator-metal transition, positive *ac* magnetoresistance, and frequency-dependent peaks in the *ac* magnetoresistance. By contrast, the CH-LMO sample crystallizes in the orthorhombic structure and is a canted antiferromagnetic insulator. Differences in magnetic and electrical properties between CH and MW samples indicate that more charge carriers ($e_g$ holes of $Mn^{4+}$ :$t_{2g}^3 e_g^0$ ions) are doped in the MW-LMO sample. Hopping of spin-polarized holes between $Mn^{4+}$ and $Mn^{3+}$ leads to ferromagnetism and metallic behavior via Zener's double exchange interaction as in the divalent cation-doped $LaMnO_3$. In addition to substituting a lower valent cation (eg. $Sr^{2+}$) for the higher valent rare earth ion $La^{3+}$ to dope holes, there are two other processes by which charge carriers can be doped in $LaMnO_3$. One method is creation of cation vacancies as in $La_{1-x}Mn_{1-x'}O_3$ through synthesis of samples at lower temperatures ($T$ = 700-1000 °C) by a soft chemistry route (sol-gel, citrate, co-precipitation, etc.) [22,23,24]. This cation-deficient compound is often denoted as $LaMnO_{3+d}$. Since excess oxygen can't be accommodated in the perovskite structure, overall charge neutrality is achieved by adjusting cation valencies. While La vacancies up to 15-18 % support long-range ferromagnetism and metallic behavior of resistivity, more than 5 % vacancies at the Mn site degrade long-range ferromagnetism and result in insulating behavior since the path for charge transfer is severely disrupted [**Error! Bookmark not defined.**,**Error! Bookmark not defined.**,**Error! Bookmark not defined.**,25,26,27,28,29]. Another possibility is suppression of the cooperative Jahn-Teller distortion (CJT) associated with $Mn^{3+}$ ($d^4$: $t_{2g}^3e_g^1$) ions by microwave irradiation. The suppression of CJT can proceed through oxidation of a fraction of $Mn^{3+}$ into $Mn^{4+}$ ions via changes in Mn-O bond lengths, Mn-O bond angle, and lattice constants. Microwave irradiation-induced suppression of CJT was recently reported in the insulating spinel $LiMn_2O_4$ [30] and $CuFe_2O_4$ [31]. Detailed X-ray spectroscopy studies are needed to probe local structural changes in the MW sample, but this is beyond the main aim of this work. Let us focus on the high-frequency magnetoresistance.

The positive sign of *ac* magnetoresistance observed in the MW-LMO sample with increasing frequency cannot be understood within the framework of the Zener's double exchange mechanism, which allows negative magnetoresistance due to suppression of spin disorder scattering under a magnetic field. A likely possibility is that dynamic current in the sample creates a transverse *ac* magnetic field (Oersted field) and this *ac* magnetic field sets the magnetization of the sample to oscillate. As the strength of the *dc* magnetic field increases, the spin system (combination of $Mn^{3+}$-$Mn^{4+}$ pairs) of the sample will resonantly absorb maximum MW power from the *ac* magnetic field when $H_{dc} = H_r$, ($H_r$ is the resonance field), leading to precession of magnetization about the direction of the longitudinal *dc* magnetic field. This is the current-driven electron spin resonance. As the frequency increases, the surface impedance of the sample increases as the current tends to flow in a shell of thickness, δ, the skin depth. The surface impedance of a metallic conductor is $Z = \sqrt{\omega\rho}[\sqrt{\mu_R} + i\sqrt{\mu_L}]$, where the effective resistive permeability is $\mu_R = |\mu_t| + \mu_t^{''}$ and the inductive permeability is $\mu_L = |\mu_t| - \mu_t^{''}$, where $|\mu_t|$ is the modulus of the complex transverse permeability [32,33]. Thus, the real part of the impedance is dependent on the modulus of the complex permeability and magnetic loss $\mu_t^{''}$. At resonance, $\mu_t^{''}$ will have its maximum value (and $\mu_t'$ will cross zero), leading to a peak in the observed *ac* resistance.

The observed feature in the field dependence of the *ac* magnetoresistance has a resemblance to the field profile of *dc* voltage induced by the inverse spin Hall effect (ISHE) in FMM/NM bilayers, where NM is a non-magnetic metal possessing strong spin-orbit coupling such as Pt or W and FMM is a ferromagnetic metal. The *dc* voltage measured in this heterostructure goes through a peak value when the magnetization of FMM layer is driven into ferromagnetic resonance by an *rf* magnetic field. The later is achieved by inserting the bilayer sample either inside a microwave cavity and exposing it to the maximum amplitude of magnetic field component microwave electromagnetic field [34,35] or placing the bilayer on a coplanar waveguide, where alternating current flowing in the signal line creates a microwave magnetic field that inductively couples to the magnetization of FM layer and induces precession when resonance condition is met while sweeping the external *dc* magnetic field in transverse direction [36]. Precession of the magnetization pumps pure spin current into the NM where it is transformed into a charge current by spin-orbit interaction and detected as a *dc* voltage. A heavy metal is not used in our method and the *ac* current flows directly to the sample. In our experiment, an impedance (ac resistance and reactance) is measured instead of a *dc* voltage. We are unaware of a published work on the observation of ISHE in a bilayer of paramagnetic and Pt.

We analyze the field dependence of *MR* data using two Lorentzian functions similar to the analysis of *dc* voltage in inverse spin Hall effect measurements.

$$MR(\%) = A_{sym}\frac{(\Delta H)^2}{(H-H_r)^2+(\Delta H)^2} + B_{asym}\frac{(\Delta H)(H-H_{res})}{(H-H_r)^2+(\Delta H)^2} + C \quad (1)$$

where $A_{sym}$ and $B_{asym}$ are the coefficients of symmetric and antisymmetric Lorentzian functions, $H_r$ is the resonance field, $\Delta H$ is the full line width at half-maximum and $C$ is an offset fitting parameter. The fit to Eq. (1) is shown in Fig. 5(a). The $f$ vs $H_r$ data are found to have a linear dependence described by

$$f = \frac{\gamma}{2\pi} \times H_r \quad (2)$$

This is the condition for the occurrence of electron spin resonance or paramagnetic resonance, where $\gamma$ is the gyromagnetic ratio ($\gamma = g\mu_B/\hbar$). The fit yields $\gamma/2\pi$ = 2.816 MHz/Oe, from which we obtain the *g*-factor = 2.012, which is close to $g = 2$ for free electrons.

Finally, to verify that the observed features in the *ac* magneto-transport are due to paramagnetic resonance, we carried out another independent measurement in which the sample is placed on the signal line of a co-planar waveguide (CPW) and the change in microwave transmission through the CPW is monitored when an externally applied *dc* magnetic field is swept. The $S_{21}$ parameter which measures the transmission of electromagnetic waves in the CPW can show a dip when the sample absorbs power from an MW magnetic field arising from alternating current flowing in the signal line of the CPW. Fig. 6(a) shows $S_{21}$ vs $H$ at selected frequencies between 1 and 7 GHz at 300 K, while Fig. 6(b) shows the 2D contour plot of the $S_{21}$ parameter measured from 10 MHz to 7.8 GHz. The darker shade of the contour represents the dip. As the field was swept, the $S_{21}$ parameter shows a dip at $H = \pm H_r$ and the dip occurs at a higher $H$ for a higher $f$. We fitted the data using eq. 1, with $S_{21}$ replacing $MR$. The obtained $H_r$ against the frequency exhibits linear behavior. Using eq. 2, we obtain $\gamma/2\pi = 2.813$ MHz/Oe and the $g$-value = 2.010, which agrees very closely with the values obtained in the magnetoresistance experiment. Hence, the peak feature observed in the high-frequency magnetoresistance is most likely related to the power absorption from the magnetic field due to paramagnetic resonance.

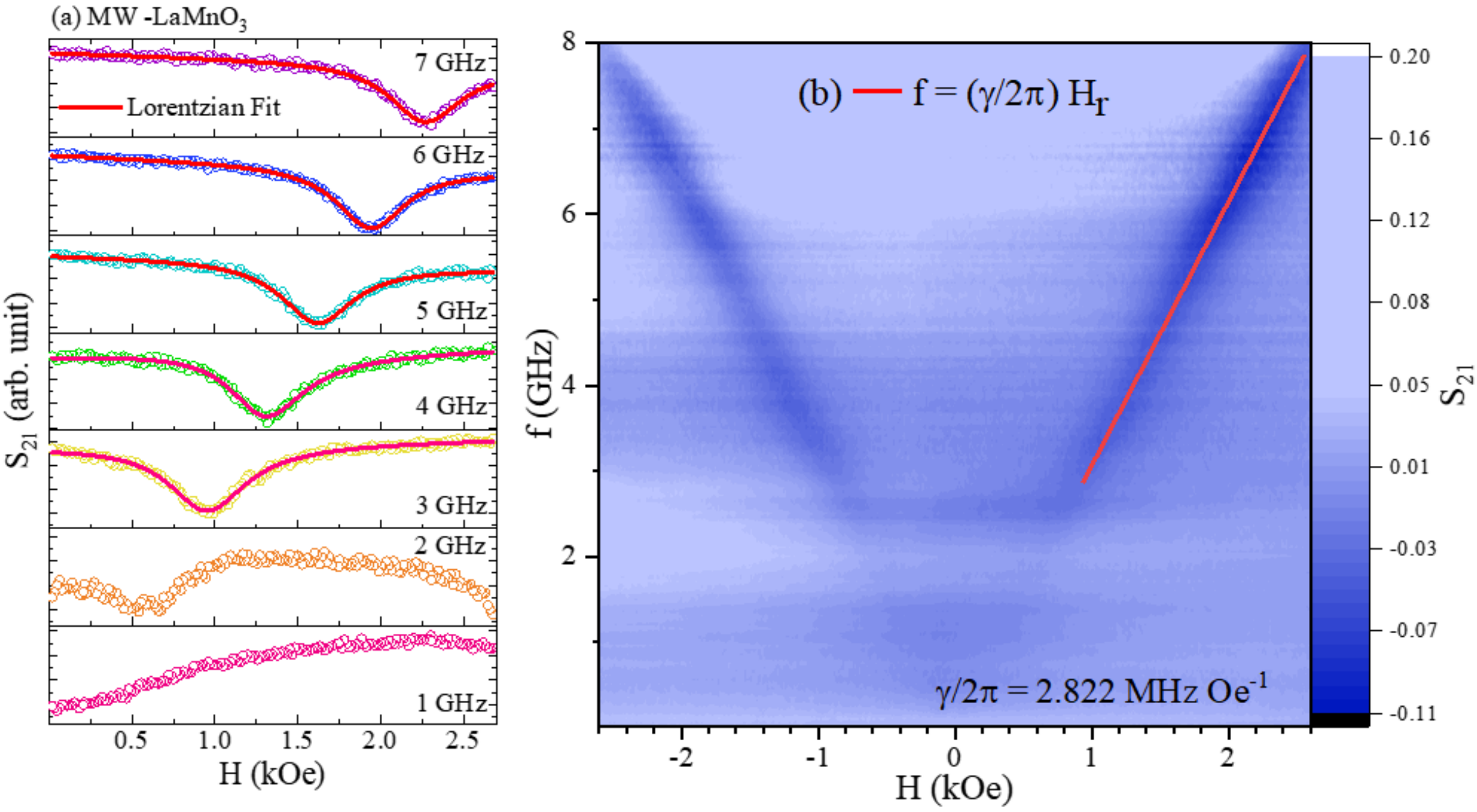

**Figure 6. (a)** $S_{21}$ parameters measured with N5230A VNA, at selected frequencies from $f = 1$ to 7 GHz at 300 K. The deep shifts towards higher field with increasing frequencies. **(b)** Contour plots of $S_{21}$ parameter. The darker shade depicts the dips, which are observed to shift toward higher fields as frequency increases.

## V. CONCLUSIONS

In summary, $LaMnO_3$ samples were synthesized by microwave (MW) irradiation heating and conventional heating (CH) separately, and their properties were compared. The MW-$LaMnO_3$ sample exhibited higher magnetization, ferromagnetic Curie temperature, and much lower resistivity than the CH-LMO sample. The magnetic-field dependence of high-frequency magnetoresistance in the MW-LMO sample showed an anomalous positive peak at a critical field when $f$ = 1.4 GHz, and this critical field increased linearly with the frequency of the current. The origin of high-frequency magnetoresistance in the MW-$LaMnO_3$ sample was suggested to be paramagnetic resonance driven by the alternating magnetic-field component of the high-frequency current. Its absence in the CH-LMO indicates that a certain fraction of itinerant carriers (holes) may be needed to observe this effect using this impedance method.

**Acknowledgments**: R. M thanks the Ministry of Education, Singapore, for supporting this work (Grant numbers: A-8000462-00-00, A-8001933-00-00 and A-8000924-00-00)

**Data availability**

The data that support the findings of this study are available from the corresponding author upon reasonable request.